\documentclass[prd,amsmath,amssymb,superscriptaddress,floatfix,nofootinbib,notitlepage]{revtex4-1}

\usepackage{bbold}
\usepackage{color}
\usepackage{graphicx}
\usepackage{hyperref}

\graphicspath{ {plots/} }

\newcommand{\Eq}[1]{{Eq.~({\ref{#1}})}}
\newcommand{\Eqs}[1]{{Eqs.~({\ref{#1}})}}
\newcommand{\Fig}[1]{{Fig.~{\ref{#1}}}}
\newcommand{\ave}[1]{{\langle{#1}\rangle}}
\newcommand{\bea}{\begin{eqnarray}}
\newcommand{\eea}{\end{eqnarray}}
\newcommand{\beq}{\begin{equation}}
\newcommand{\eeq}{\end{equation}}
\newcommand{\beas}{\begin{eqnarray*}}
\newcommand{\eeas}{\end{eqnarray*}}

\def\p{{\bf p}}
\def\q{{\bf q}}
\def\x{{\bf x}}

\def\Fonez{{F_1(0)}}
\def\Ftwoz{{F_2(0)}}
\def\Fones{{F_1(M_{s,0})}}
\def\Ftwos{{F_2(M_{s,0})}}

\begin{document}

\title{Inhomogeneous chiral condensates in three-flavor quark matter
} 

\author{Stefano~Carignano}
\affiliation{Departament de F\'isica Qu\`antica i Astrof\'isica and Institut de Ci\`encies del Cosmos, Universitat de Barcelona, Mart\'i i Franqu\`es 1, 08028 Barcelona, Catalonia, Spain.}

\author{Michael~Buballa}
\affiliation{Theoriezentrum, Institut f\"ur Kernphysik, Technische Universit\"at Darmstadt, 
Schlossgartenstr.\ 2, D-64289 Darmstadt, Germany}

\begin{abstract}

We investigate the effect of strange-quark degrees of freedom on the formation of inhomogeneous chiral condensates
in a three-flavor Nambu--Jona-Lasinio model 
in mean-field approximation.
A Ginzburg-Landau study complemented
by a stability analysis allows us to determine in a general way the location of the critical and Lifshitz points, together with the 
phase boundary where 
the (partially) chirally restored phase becomes unstable against developing inhomogeneities, without resorting to specific 
assumptions on the shape of the chiral condensate.  We discuss the resulting phase structure and study the influence of 
the bare strange-quark mass $m_s$ and the axial anomaly on the size and location
of the inhomogeneous phase compared to the first-order transition associated with homogeneous matter.
We find that, as a consequence of the axial anomaly, critical and Lifshitz point split. For realistic strange-quark masses
the effect is however very small and becomes sizeable only for small values of $m_s$.
\end{abstract}

\maketitle

\section{Introduction}

Mapping the phase diagram of QCD at nonvanishing temperature $T$ and quark chemical potential $\mu$ is one of the major 
goals in strong interaction physics \cite{Kumar:2013cqa,Friman:2011zzA}. Lattice-QCD calculations revealed that chiral symmetry, which is spontaneously broken in vacuum, gets approximately restored at high temperature via a smooth crossover transition
\cite{Aoki:2006we}. 
At low temperature but nonvanishing chemical potential, standard lattice Monte-Carlo techniques are not applicable. 
In this regime one therefore has to 
rely on other approaches to QCD, like Dyson-Schwinger equations \cite{Fischer:2018sdj}
or the Functional Renormalization Group (FRG) \cite{Fu:2019hdw}, or 
on effective models, like the Nambu--Jona-Lasinio (NJL) or the quark-meson model.
Assuming spatially uniform chiral condensates, these approaches typically predict a first-order phase transition
from the chirally broken to the (approximately) restored phase, ending at a critical end point
\cite{Fischer:2018sdj,Fu:2019hdw,Asakawa:1989bq,Scavenius:2000qd}. 
In the two-flavor chiral limit one finds instead a tricritical point where the first-order phase boundary at low $T$ is 
joined to a second-order one which replaces the crossover in the high-$T$--low-$\mu$ region.
For simplicity we will refer to both the tricritical point and the critical end point as `critical point' (CP) in the following.

Some time ago it was found that spatially inhomogeneous condensates could be favored over homogeneous ones
in certain regions of the phase diagram \cite{Kutschera:1989yz,Nakano:2004cd,Nickel:2009ke,Nickel:2009wj}, see
Ref.~\cite{Buballa:2014tba} for a review. In the chiral limit one then finds a so-called Lifshitz point (LP) 
where three phases, the homogeneous chirally broken phase, the restored phase and the inhomogeneous phase, meet. 
In particular it was found for the two-flavor NJL model in mean-field approximation 
that the inhomogeneous phase completely covers the first-order phase boundary between the homogeneous phases \cite{Nickel:2009wj} and that the LP exactly coincides with the CP \cite{Nickel:2009ke}.\footnote{This holds for the standard NJL model with scalar and pseudoscalar interactions. If vector interactions are added, the LP stays at the same temperature, while the CP moves to lower $T$ 
so that it is covered by the inhomogeneous phase and thus
 disappears from the phase diagram \cite{Carignano:2010ac}.}
Recently we have generalized this result to the case away from the chiral limit \cite{Buballa:2018hux}. 
A similar picture also emerged from a truncated Dyson-Schwinger approach to QCD, although the coincidence of CP and LP
has not yet been shown rigorously \cite{Muller:2013tya}.  
Indications for the presence of an inhomogeneous phase have also been found in a very recent FRG study of the QCD
phase diagram  \cite{Fu:2019hdw}. 

In the present article we want to investigate the effect of strangeness degrees of freedom on the locations of the CP 
and the LP in the NJL model.
 To this end we will employ the three-flavor NJL model and perform 
a Ginzburg-Landau (GL) analysis.
In order to get a better feeling on the form of the phase diagram, this GL study will be supplemented by a stability analysis which, 
under certain conditions, can provide us the location of the phase boundary between the inhomogeneous and the partially chirally restored phase.

Our motivation for this study is twofold: 
First, in the two-flavor model 
 the CP and the LP are typically found in a temperature and density regime where strange-quark 
effects may not be negligible. Including strangeness degrees of freedom will thus be a step toward a more realistic description.
A first study of inhomogeneous phases in three-flavor matter has been performed in {Ref.}~\cite{Moreira:2013ura}, albeit only for 
vanishing temperature and employing a simple explicit ansatz for the spatial dependence of the chiral condensate.
The authors found that the coupling of strange and non-strange quarks, which is related to the axial anomaly, enlarges the
inhomogeneous domain.  Our investigation will be complementary, as its range of validity will be close to the locations of 
the CP and the LP
at non-vanishing temperature, as well as along the phase  boundary between the inhomogeneous 
and the partially restored phase.
 Moreover, it does not rely on a certain ansatz for the inhomogeneity and allows for a more transparent
analysis of the role of the axial anomaly. 

Our second motivation is the fact that QCD with three very light quark flavors is expected to have a first-order
chiral phase transition even at $\mu = 0$ \cite{Pisarski:1983ms}. Hence if in three-flavor QCD the LP coincided with the CP, 
as it does in the two-flavor NJL model, it would mean that the inhomogeneous phase would also reach down to $\mu = 0$ for
small enough quark masses. At least in principle it could then be studied on the lattice in this case. 
As a first step we therefore want to investigate the behavior of CP and LP in the three-flavor NJL model. 
While it has been known for a long time
that the CP indeed moves to lower chemical potentials
when lowering the quark masses and eventually reaches the $T$ axis \cite{Fukushima:2008}, 
until recently the behavior of the LP has not yet been studied for the three-flavor case. In particular it has not been checked whether CP and LP still coincide in that case.\footnote{
In a recent proceedings article~\cite{Buballa:2019clw} we have presented parts of the present analysis in a simplified form
and concluded that CP and LP do not coincide. In the present paper we perform an extended study and back up the results 
by numerical examples in order to get a better feeling of the quantitative relevance of this finding.}

Our paper is organized as follows. In Sect.~\ref{sec:strange} we introduce the three-flavor NJL model and derive a
general expression for the mean-field thermodynamic potential as functional of several strange and non-strange chiral condensates
with arbitrary spatial dependence. Based on this we perform in Sect.~\ref{sec:GL} a GL  and a stability analysis, focusing on the 
locations of the CP and the LP as well as the position of the phase transition to the partially restored phase. 
In Sect.~\ref{sec:num} we give a quantitative illustration of the results by numerical 
evaluation of the GL coefficients before we draw our conclusions in Sect.~\ref{sec:conclusions}.

\section{NJL model setup}
\label{sec:strange}

Our starting point is the three-flavor NJL Lagrangian
\beq
\mathcal{L} = \bar\psi\left( i\gamma^\mu \partial_\mu - \hat m \right) \psi 
+ \mathcal{L}_4 + \mathcal{L}_6
\eeq
where $\psi = (u,d,s)^T$ denotes a quark field with three flavor degrees of freedom
and $\hat m$ is the matrix containing their current masses. 
The last two terms describe a $U(3)_L\times U(3)_R$ invariant four-point interaction
\beq
\mathcal{L}_4 
=  G \sum_{a=0}^8\left[ (\bar\psi\tau_a\psi)^2 + (\bar\psi i\gamma_5 \tau_a \psi)^2 \right],
\eeq
and a six-point (``Kobayashi - Maskawa - 't Hooft'', KMT) interaction 
\beq
\mathcal{L}_6 = -K \left[ \mathrm{det}_f \bar\psi(1+\gamma_5)\psi +  \mathrm{det}_f  \bar\psi(1-\gamma_5)\psi \right] \,,
\eeq
which is $SU(3)_L\times SU(3)_R$ symmetric but breaks the $U(1)_A$ symmetry,
thus mimicking the axial anomaly.
In the former $\tau_a$, $a=1,\dots,8$, denote Gell-Mann matrices in flavor space while 
$\tau_0 = \sqrt{2/3} \,\mathbb{1}$ is proportional to the unit matrix.

Following standard procedures, we perform a mean-field approximation, considering  
scalar and pseudoscalar flavor-diagonal condensates, 
\beq
   \sigma_f(\x)  \equiv  \ave{\bar f f}(\x)  \,, \qquad   \pi_f(\x)  \equiv   \ave{\bar f\, i \gamma^5 f}(\x) \,, \qquad f = \{u,d,s\} \,,
 \eeq
allowing them to be spatially dependent. The mean-field quark self-energies can then be expressed in terms 
of the mass operators 
       \beq
       {\hat M}_f(\x) = m_f - 4G \sigma_f(\x) + 2K \Big(\sigma_g(\x)\sigma_h(\x) - \pi_g(\x)\pi_h(\x) \Big) + i \gamma^5 \Big[ -4G \pi_f(\x) - 2K(\sigma_g(\x)\pi_h(\x) + \pi_g(\x) \sigma_h(\x)) \Big] \,,
       \label{eq:Mf}
       \eeq
where $ f \neq g \neq h \neq f$.       
We furthermore assume isospin symmetry in the light sector, $m_u = m_d \equiv m_\ell$, and thus
\beq
       \sigma_u(\x) =  \sigma_d(\x) \equiv \sigma_\ell(\x)
\eeq       
due to the flavor symmetry of the interaction.
For the pseudoscalar condensates, given the isospin structure in the light sector, we identify 
\beq
       \pi_u(\x) = -\pi_d(\x) \equiv \pi_\ell(\x),
\eeq 
while we neglect the pseudoscalar condensate in the strange sector, $\pi_s = 0$.   
The latter is motivated by the fact that pseudoscalar condensates are usually suppressed by the presence of 
non-vanishing bare quark masses, which we will always consider in the strange sector.     

The mass operators then reduce to 
\begin{alignat}{1}
{\hat M}_u(\x) &= m_\ell - \left[4G -2K\sigma_s(\x)\right] \left(\sigma_\ell(\x) + i\gamma^5 \pi_\ell(\x)\right) \,,
\nonumber \\
{\hat M}_d(\x) &= m_\ell - \left[4G -2K\sigma_s(\x)\right] \left(\sigma_\ell(\x) - i\gamma^5 \pi_\ell(\x)\right) \,,
\nonumber \\
 {\hat M}_s(\x) &= m_s - 4G\sigma_s(\x)   + 2K\left(\sigma_\ell^2(\x) + \pi_\ell^2(\x)\right)  \,,
\label{eq:Mells}
\end{alignat}    
which is a considerable simplification compared to  \Eq{eq:Mf}  but still catches the most important structures.
For instance, it is compatible with the ansatz of Ref.~\cite{Moreira:2013ura}, which we will briefly discuss at the end of this section.

Neglecting fluctuations, the mean-field thermodynamic potential at temperature $T$ and quark chemical potential $\mu$
is then given by 
\beq
\Omega(T,\mu) = -\frac{T}{V} \, \mathbf{Tr}\, \log \frac{S^{-1}}{{T}} \,
+\, \frac{1}{V}\int\limits_V d^3x \, \Big\{ 2G \big[2(\sigma_\ell^2 + \pi_\ell^2) + \sigma_s^2 \big] -4K (\sigma_\ell^2 + \pi_\ell^2) \sigma_s \Big\} \,,
\label{eq:Omega3}
\eeq
where $V$ is a quantization volume and $\mathbf{Tr}$ denotes a functional trace to be taken in the Euclidean space-time 
($(it,\x) \in V_4 = [0,\frac{1}{T}] \times V$)
and over internal (Dirac, color, and flavor) degrees of freedom.
The inverse dressed quark propagator is given by
$S^{-1} = \mathrm{diag}_f(S^{-1}_u,S^{-1}_d,S^{-1}_s)$
with the flavor components
\beq
       S^{-1}_f(x) =  i\gamma^\mu \partial_\mu + \mu\gamma^0 - {\hat M}_f(\mathbf{x})  \,. 
\label{eq:Sinv}       
\eeq
From the above expressions we can see that the different flavors only couple through the six-point interaction 
and thus for $K=0$ we obtain the usual expression for the two-flavor model plus a decoupled s-quark.

Having set up the model, we can now analyze its phase structure 
by minimizing $\Omega(T,\mu)$ with respect to the condensates.
When restricting the analysis to homogeneous condensates, this is straightforward and has been done many times
in the literature, see, e.g., Ref.~\cite{Buballa:2005}.  
With realistic parameters one basically obtains the result already mentioned in the Introduction,
i.e., a first-order phase transition at low temperatures
turning into a crossover at higher $T$, both of which can mainly be associated with the chiral restoration in the
light-flavor sector.  Further outside, i.e., at larger values of $\mu$ or $T$, there is another transition (usually a crossover)
related to
the partial symmetry restoration in the strange-quark sector. Although of minor importance for our GL analysis,
this second transition will play a role for the behavior of the inhomogeneous phase at higher chemical potentials, as we 
will briefly discuss at the end of section \ref{sec:num}.

The numerical analysis of the model phase diagram allowing for inhomogeneous condensates is obviously much more 
challenging, as one should in principle allow for arbitrary spatial modulations of  $\sigma_\ell$, $\pi_\ell$, and $\sigma_s$.
The authors of Ref.~\cite{Moreira:2013ura} have therefore restricted themselves to a relatively simple ansatz,
namely a chiral density wave in the light sector, 
$\sigma_\ell(\x) = \sigma_0 \cos(\q\cdot\x)$,  $\pi_\ell(\x) = \sigma_0 \sin(\q\cdot\x)$, $m_\ell = 0$,        
embedded into a homogeneous strange-quark background $\sigma_s = \mathit{const.}$
Inserting this into \Eq{eq:Mells}, one finds that $\hat M_s = m_s - 4G\sigma_s   + 2K\sigma_0^2$ is homogeneous as well,
while $\hat M_u$ and $\hat M_d$ form an isospin doublet $\hat M_\ell = M_0 \exp(i\gamma^5 \tau^3 \q\cdot\x)$
with amplitude $M_0 = -(4G -2K\sigma_s) \sigma_0$.
This feature allowed the authors to find the eigenvalue spectrum of the corresponding Hamiltonian by adopting
the known two-flavor result~\cite{Dautry:1979}.
It should be noted, however, that this straightforward generalization of a known two-flavor solution was only possible
because the space dependence of the $\sigma_\ell$ and $\pi_\ell$ contributions to $\hat M_s$ cancel each other for this
ansatz. 
In contrast, the purely scalar  ``real-kink crystal'', which in the two-flavor model is energetically favored over the
chiral density wave \cite{Nickel:2009wj}, cannot selfconsistently be generalized in this simple way since coupling 
it to a homogeneous $\sigma_s$ would still lead to an oscillating $\hat M_s$ as long as $K$ is nonzero.

In the remainder of this work we will not resort to a particular ansatz for the condensates 
but consider general functions $\sigma_\ell(\x)$, $\pi_\ell(\x)$, and $\sigma_s(\x)$ 
both within a GL as well as a stability analysis of the homogeneous phase.
This will allow us, under certain conditions,
to locate the position of the Lifshitz point  together with the critical point and
the phase boundary where inhomogeneous condensates become energetically favored.

\section{Ginzburg-Landau and stability analysis}
\label{sec:GL}
 
The GL expansion of the thermodynamic potential allows for a systematic study of 
the phase structure
while requiring only a limited number of assumptions.
Generally, it corresponds to an expansion of the free energy in terms of one or more order parameters and their 
gradients, assuming that these quantities are small. 
In the context of chiral symmetry breaking the order parameter is usually identified with the chiral condensate, 
and the expansion is performed about the chirally restored solution where the condensate vanishes.
Of course, this only works in the chiral limit, while in the presence of nonzero bare quark masses there is no
chirally restored solution and therefore the expansion has to be performed around nonvanishing condensate 
values~\cite{Buballa:2018hux}.  

On the other hand, in a realistic calculation the bare strange-quark mass is much larger than the  bare masses of the up and down quarks. 
We may therefore study a partially simplified problem where we take into account $m_s\neq 0$ but consider $m_\ell=0$. 
In this case a two-flavor restored solution $\sigma_\ell =\pi_\ell= 0$ exists, which we take as the expansion point of the GL analysis.
In the strange-quark sector we proceed similarly as in \cite{Buballa:2018hux} for the case of light quarks away from the chiral limit and expand around a 
$T$ and $\mu$-dependent but spatially constant  condensate $\sigma_s^{(0)}$, corresponding to a stationary point of $\Omega$
at $\sigma_\ell = \pi_\ell = 0$. To this end we write
\beq
       \sigma_s(\x) = \sigma_s^{(0)} + \delta\sigma_s(\x)
\label{eq:sigmas}
\eeq
and introduce the order parameters
\beq
       \Delta_\ell(\x) = -4G \left(\sigma_\ell(\x) + i\pi_\ell(\x)\right)       
       \quad \text{and}\quad 
       \Delta_s(\x) = -4G\,\delta\sigma_s(\x) \,,
\label{eq:Deltaf}
\eeq
which are proportional to the fluctuations around our expansion point and have the dimension of a mass. 
We then write
\beq
       \Omega[\sigma_\ell,\pi_\ell, \sigma_s]
        =  \Omega[0,0,\sigma_s^{(0)}] +  \frac{1}{V} \int d^3x \, \omega_{GL}(\Delta_\ell,\Delta_s) \,,
\eeq
and expand $\omega_{GL}$ in powers of  $\Delta_\ell$ and $\Delta_s$ and gradients thereof.

Note that in \Eq{eq:Deltaf}
we combined the light-quark scalar and pseudoscalar condensates to a complex order parameter.
In fact, as a consequence of the remaining two-flavor chiral symmetry, $\sigma_\ell$
and $\pi_\ell$ can be rotated into each other by a symmetry transformation and therefore $\omega_{GL}$
can only depend on the modulus of $\Delta_\ell$ but not on its phase.
Moreover, since we expand about the two-flavor restored phase, only even powers of $|\Delta_\ell|$ are allowed.
For $\Delta_s$, on the other hand, we can also have odd terms. 
Finally, the possible gradient terms are restricted by the requirement that the potential must be a scalar under spatial rotations.

Treating initially both light and strange order parameters as well as gradients to be of the same order,
the GL functional up to order 4 thus takes the form
\begin{alignat}{1}
\omega_{GL}(\Delta_\ell,\Delta_s)
= \phantom{+} &\alpha_2 |\Delta_\ell|^2 + \alpha_{4,a} |\Delta_\ell|^4 + \alpha_{4,b} |\nabla\Delta_\ell|^2 + \dots
\nonumber\\
+ &\beta_1 \Delta_s  + \beta_2 \Delta_s^2  + \beta_3 \Delta_s^3 + \beta_{4,a} \Delta_s^4 + \beta_{4,b} (\nabla\Delta_s)^2
+ \dots
\nonumber\\
+ &\gamma_3 |\Delta_\ell|^2 \Delta_s + \gamma_4 |\Delta_\ell|^2 \Delta_s^2  + \dots \,,
\label{eq:Omega_GL_s}
\end{alignat}
where the $\alpha$- and $\beta$-terms correspond to the contributions from the light and strange condensates, respectively, 
and the $\gamma$ terms describe the mixing.

Since we are expanding about a homogeneous stationary point, the linear coefficient $\beta_1$ associated with $\Delta_s$ has to vanish,
\beq
       \left. \frac{\partial \omega_{GL}}{\partial\Delta_s}\right|_{\Delta_\ell=\Delta_s=0} = \beta_1 = 0 \,.
\eeq
This condition is of course equivalent to a gap equation for
$\sigma_s^{(0)}$ at given $T$ and $\mu$. 

For vanishing $\gamma_i$  the GL analysis of the non-strange sector would be identical to the 
two-flavor case in the chiral limit, which has been analyzed in Ref.~\cite{Nickel:2009ke}.
More specifically, the location of the CP, where the first-order phase transition
turns into a second-order one, is given by the point 
where both the quadratic and quartic coefficients $\alpha_2$  and  $\alpha_{4,a}$ vanish,
while the LP appears where both $\alpha_2$ and the lowest-order gradient coefficient $\alpha_{4,b}$ become 
zero.  
For the standard two-flavor NJL model one finds that $\alpha_{4,a} = \alpha_{4,b}$ and as a consequence the CP and the LP 
coincide \cite{Nickel:2009ke}.

In order to study how these relations get modified in the presence of a third flavor coupled to the light sector, as reflected by $\gamma_i \neq 0$, 
the first step is to eliminate $\Delta_s$ by extremizing the thermodynamic potential with respect to this function.
To this end we employ the Euler-Lagrange equations,
\beq
      \frac{\partial \omega_{GL}}{\partial \Delta_s} -\partial_i \frac{\partial \omega_{GL}}{\partial \partial_i\Delta_s} = 0\,,
\eeq
which, after using the gap equation $\beta_1=0$, yields
\beq
       \Delta_s = -\frac{\gamma_3}{2\beta_2} |\Delta_\ell|^2 + \mathcal{O}(\Delta_\ell^4)  \equiv   \Delta_s^\mathit{extr}\,.
\label{eq:Dselim}       
\eeq
Here, the orders in $\Delta_\ell$ and gradients are treated equally,  
$\mathcal{O}(\nabla^n)=\mathcal{O}(\Delta_\ell^n)$, as we already did when we wrote down \Eq{eq:Omega_GL_s}. 
From \Eq{eq:Dselim} we then see, however, that $\Delta_s$ is of the order $\mathcal{O}(\Delta_\ell^2)$.\footnote{ 
This is reminiscent of the 
case when vector interactions are included in the model: by solving the equation of motion for the vector condensate, one also
finds it to be quadratic in the amplitude of the mass amplitude \cite{Carignano:2010ac,Carignano:2018hvn}.}
Inserting \Eq{eq:Dselim} back into \Eq{eq:Omega_GL_s} and keeping only terms up to the order $\mathcal{O}(\Delta_\ell^4)$,
one then obtains
\beq
       \omega_{GL}(\Delta_\ell,\Delta_s^\mathit{extr}) 
       = \alpha_2 |\Delta_\ell|^2 + \left(\alpha_{4,a} - \frac{\gamma_3^2}{4\beta_2} \right) |\Delta_\ell|^4 
       + \alpha_{4,b} |\nabla\Delta_\ell|^2 + \dots
\eeq
We thus find that the quartic term in $\Delta_\ell$ gets an additional contribution through the coupling to the strange quarks,
while the gradient term does not. 
So the relevant equations for locating the CP and the LP become 
\beq
      \alpha_2 =  0 \;\; \text{and} \;\; \alpha_{4,CP} \equiv \alpha_{4,a}  - \frac{\gamma_3^2}{4\beta_2} = 0  \quad \text{at the CP,} \qquad 
      \alpha_2 = 0 \;\; \text{and} \;\;  \alpha_{4,LP} \equiv  \alpha_{4,b} = 0 \quad \text{at the LP}\,,
\label{eq:TCPLP3}
\eeq      
and therefore, even if $\alpha_{4,a}$ and $\alpha_{4,b}$ were still equal (as they are in the two-flavor model)
the CP and the LP would no longer coincide for $\gamma_3\neq 0$.

\subsection{GL coefficients}
\label{sec:GLcoeffs}
In order to make more quantitative statements on the relative position of CP and LP, we need to determine the GL coefficients explicitly.
To this end we basically follow the procedure of Ref.~\cite{Nickel:2009ke}, although the present case is technically somewhat more 
involved due to the nonvanishing strange-quark mass and the presence of the KMT term. 

In a first step we decompose the mass operator for flavor $f$ into a constant and a fluctuating piece,
\beq
       \hat M_f(\x) = M_{f,0} + \delta\hat M_f(\x)\,,
\eeq
by inserting \Eqs{eq:sigmas} and (\ref{eq:Deltaf}) into \Eq{eq:Mells}. Explicitly one finds
\begin{alignat}{2}
       &M_{s,0} = m_s - 4G\sigma_s^{(0)}\,, \quad &
       &\delta\hat M_s(\x)  = \Delta_s(\x) + {\kappa} |\Delta_\ell|^2\,,
\\
       &M_{u,0}=M_{d,0} = 0\,, \quad &
       &\delta\hat M_{u\atop d}(\x) = \left[1 + {\delta + \kappa}\Delta_s(\x) \right]
       \left(\mathrm{Re} \Delta_\ell(\x) \pm i\gamma^5 \mathrm{Im} \Delta_\ell(\x)\right)  \,,
\end{alignat}
where we have introduced $\kappa = K/8G^2$ and $\delta = \kappa(M_{s,0}-m_s)$.
The mean-field thermodynamic potential \Eq{eq:Omega3} can then be written as
\beq
\Omega = -N_c \sum_f \frac{T}{V} \, \mathrm{Tr}_{D,V_4} \log(S_{f,0}^{-1}-\delta \hat M_f)
                 \,+\,  \Omega_\mathit{cond} \,,
\label{eq:OmegadelM}
\eeq
where $S_{f,0}^{-1}(x) = i\gamma^\mu \partial_\mu + \mu\gamma^0 - M_{f,0}$ is the inverse bare propagator 
of a free fermion with mass $M_{f,0}$.  We have already turned out the trivial color trace 
(leading to a factor of $N_c$) and write the flavor trace explicitly as a sum, so that the remaining
functional trace is only to be taken in Euclidean space-time and Dirac space.

The condensate part, corresponding to the second term on the right-hand side of  \Eq{eq:Omega3}, becomes
\beq
       \Omega_\mathit{cond} = \frac{(M_{s,0}-m_s)^2}{8G} 
       + \frac{1}{V} \int\limits_V d^3x\,\omega_{GL}^\mathit{cond}(\Delta_\ell,\Delta_s) \,,
\eeq
with
\beq
       \omega_{GL}^\mathit{cond}(\Delta_\ell,\Delta_s) 
       = \frac{1}{4G} \left\{ (M_{s,0}-m_s) \Delta_s
       + ( 1 + 2\delta) |\Delta_\ell|^2 
       +  \frac{1}{2} \Delta_s^2
       + 2\kappa |\Delta_\ell|^2 \Delta_s\right\} \,,     
\eeq
from which we can straightforwardly read off the condensate contributions to the GL coefficients
$\beta_1$, $\alpha_2$, $\beta_2$, and $\gamma_3$.

In order to obtain the contributions from the `$\mathrm{Tr}\,\log$' term in \Eq{eq:OmegadelM} 
as well, we expand the logarithm into a Taylor series about $S_0^{-1}$:
\beq
       \mathrm{Tr}_{D,V_4} \log(S_{f,0}^{-1}-\delta \hat M_f)
       =
        \mathrm{Tr}_{D,V_4} \log(S_{f,0}^{-1}) 
        - \frac{1}{n}\sum\limits_{n=1}^\infty \mathrm{Tr}_{D,V_4}  \big(S_{f,0}\, \delta  \hat M_f\big)^n   \,.
\eeq
The functional traces are given by
\beq
\mathrm{Tr}_{D,V_4} \big( S_{f,0} \, \delta \hat M_f \big)^n =
       \int \prod_{i=1}^n d^4x_i  \, \mathrm{tr}_\mathrm{D} \left[
       S_{f,0}(x_n,x_1) \delta \hat M_f(\x_1)S_{f,0}(x_1,x_2) \delta \hat M_f(\x_2) \dots S_{f,0}(x_{n-1},x_n) \delta \hat M_f(\x_n)\right]\,,
\label{eq: TrDV}
\eeq
where the integrals are again over $V_4$, and $\mathrm{tr}_\mathrm{D}$ indicates a trace in Dirac space.
Noting that $S_{f,0}$ is the standard propagator of a free fermion with mass $M_{f,0}$ at 
temperature $T$ and chemical potential $\mu$,
the evaluation of the Dirac trace is straightforward using the momentum-space representation of $S_{f,0}$
in the Matsubara formalism.
After performing a gradient expansion of $\delta \hat M_f(\x_i)$ about $\x_1$, i.e.,
$\delta \hat M_f(\x_i) = \delta \hat M_f(\x_1) + \nabla \delta \hat M_f(\x_1)\cdot(\x_i - \x_1) + \dots$,
one can also perform the integrations over all space-time variables $x_i \neq x_1$
and then compare the results with \Eq{eq:Omega_GL_s} to read off the GL coefficients.
Here some attention has to be payed to the fact that, as a consequence of the KMT coupling, 
$\delta  \hat M_f$ contains linear and quadratic terms in the fluctuations $\Delta_i$
and therefore $(S_{f,0}\, \delta  \hat M_f)^n$ contains different orders as well.
We find
\begin{alignat}{1}
\alpha_2 & = (1+2\delta)\frac{1}{4G} + \frac{1}{2}(1+\delta)^2  \Fonez +  \frac{1}{2}\kappa M_{s,0} \Fones\,,
\\
\alpha_{4,a} & = \frac{1}{4}(1+\delta)^4  \Ftwoz +  \frac{1}{4}\kappa^2 \left(\Fones   +2M_{s,0}^2 \Ftwos\right)\,,
\\
\alpha_{4,b} & =   \frac{1}{4} (1+\delta)^2\Ftwoz \,, \label{eq:a4b}
\\
\beta_1 & = \frac{M_{s,0}-m_s}{4G} + \frac{1}{2}M_{s,0}\Fones\,,
\\
\beta_2 & = \frac{1}{8G} + \frac{1}{4} \Fones+  \frac{1}{2} M_{s,0}^2 \Ftwos\,,
\\
\gamma_3 & = \kappa \left\{ \frac{1}{2G} + (1+\delta)  \Fonez 
                            + \frac{1}{2} \Big( \Fones +  2 M_{s,0}^2 \Ftwos \Big) \right\}\,,                                                       
\end{alignat}
where we have introduced the functions
\beq
       F_n(M) = 8 N_c \int \frac{d^3p}{(2\pi)^3}\, T\sum\limits_j \frac{1}{[(i\omega_j +\mu)^2 - \p^2 -M^2]^{n}}      \,,
\label{eq:Fn1}
\eeq
with fermionic Matsubara frequencies $\omega_j = (2j+1)\pi T$.

From the stationary condition $\beta_1 = 0$ we thus obtain
\beq
        M_{s,0}  = m_s   - 2G M_{s,0}\Fones\ \,,
\label{eq:gap}        
\eeq
which corresponds to the gap equation for a constant dressed strange-quark mass at temperature $T$ and chemical potential $\mu$
in the absence of light condensates, as expected. 
With the help of this equation some of the GL coefficients can be simplified. In particular we find
\begin{alignat}{1}
\alpha_2 & = (1+\delta) \left[\frac{1}{4G} + \frac{1}{2}(1+\delta)  \Fonez \right] \,,
\\
\beta_2 & = \frac{1}{8G} \frac{m_s}{M_{s,0}}+  \frac{1}{2} M_{s,0}^2 \Ftwos\,,
\\
\gamma_3 & = \kappa \left\{ \frac{1}{4G}\left(1 + \frac{m_s}{M_{s,0}}\right) + (1+\delta)  \Fonez 
                            +  M_{s,0}^2 \Ftwos  \right\}\,.                                                       
\end{alignat}

As we have seen earlier, the different flavors are only coupled through the six-point interaction. 
Indeed, for $K=0$ and thus $\kappa=\delta =0$, the flavor mixing coefficient $\gamma_3$ vanishes,
and the $\alpha$ coefficients  reduce to the corresponding two-flavor expressions in the chiral limit. 
In particular, we reproduce again the result of Ref.~\cite{Nickel:2009ke} that CP and LP coincide in this case. 

For $K\neq0$, on the other hand, $\gamma_3\neq0$ and therefore the  CP and the LP would not even coincide 
if $\alpha_{4,a}$ and  $\alpha_{4,b}$ were equal,  as seen in \Eq{eq:TCPLP3}. 
Moreover, $\alpha_{4,a}$ and  $\alpha_{4,b}$ are {\it not} equal, and the two effects do not cancel each other. 
We thus find that CP and LP split for $K\neq 0$, i.e., as a consequence of the axial anomaly.

\subsection{Stability analysis}
\label{sec:stab}

The GL analysis  is valid in regimes where both the amplitudes and the gradients of the 
order parameters are small, i.e., in the vicinity of second-order phase boundaries between homogeneous phases 
or in the inhomogeneous phase close to the LP.
It is possible however to determine the entire second-order boundary between a homogeneous and an inhomogeneous 
phase as well if we drop the assumption of small gradients while keeping the assumption of small 
amplitudes \cite{Nakano:2004cd,Buballa:2018hux}.
To this end we start from \Eq{eq: TrDV} but instead of performing a gradient expansion of the mass functions
we allow the fluctuating condensates to be arbitrary periodic functions and decompose them into their
Fourier modes, i.e., we write
\beq
        \sigma_\ell(\x) = \sum\limits_{\q_k} \sigma_{\ell,\q_k} \, e^{i\q_k\cdot \x}, \quad
        \pi_\ell(\x) = \sum\limits_{\q_k} \pi_{\ell,\q_k} \, e^{i\q_k\cdot \x}, \quad
        \delta\sigma_s(\x) = \sum\limits_{\q_k} \delta\sigma_{s,\q_k} \, e^{i\q_k\cdot \x},         
\eeq
with $\q_k$ belonging to the reciprocal lattice corresponding to the periodic structure. 
Requiring real- valued condensates in coordinate space, the Fourier components for opposite momenta are related as 
$\sigma_{\ell,-\q_k} = \sigma_{\ell,\q_k}^*$, etc. 
Using again the momentum-space representations of the propagators $S_{f,0}$ as well, the integrals in  \Eq{eq: TrDV}
are readily performed. 
We can then decompose the thermodynamic potential as
\beq
       \Omega = \sum\limits_{n=0} \Omega^{(n)} \,,
\eeq
where $\Omega^{(n)}$ is of $n$th order in the fluctuations.
The linear term $\Omega^{(1)}$ is proportional to $\beta_1$ and thus vanishes by the gap equation (\ref{eq:gap}).\footnote{
More precisely we find that $ \Omega^{(1)} \propto  \beta_1\delta\sigma_{s,\q=\mathbf{0}}$ , i.e.,  it is affected only by
spatially constant modes $\q=\mathbf{0}$. Since by assumption our expansion point corresponds to a stationary homogeneous
solution of the thermodynamic potential, we can thus turn the argument around and conclude that $\beta_1$ has to vanish. 
}
For the quadratic term one finds
\beq
       \Omega^{(2)} = 8G^2 \sum\limits_k 
\Big\{ \Gamma^{-1}_\ell(\q_k^2) \left( |\delta\sigma_{\ell,\q_k}|^2 + |\delta\pi_{\ell,\q_k}|^2 \right)
       + \Gamma^{-1}_s(\q_k^2) |\delta\sigma_{s,\q_k}|^2 \big\}\,,
\label{eq:Omega2}
\eeq
with 
\begin{alignat}{1}
       \Gamma^{-1}_\ell(\q^2) 
       &= (1+\delta)\, \left[
       \frac{1}{2G} + (1+\delta)\left( \Fonez - \frac{1}{2} \q^2 L_2(\q^2;0) \right) \right],
\label{eq:Gammal}       
\\
       \Gamma^{-1}_s(\q^2) 
       &= \frac{1}{4G}\frac{m_s}{M_{s,0}}  - \frac{1}{4} \left(\q^2 + 4M_{s,0}^2\right) L_2(\q^2;M_{s,0}),     
\end{alignat}
the function
\beq
       L_2(\q^2;M) = -8 N_c \int \frac{d^3p}{(2\pi)^3}\, T\sum\limits_j 
       \frac{1}{[(i\omega_j +\mu)^2 - \p^2- M^2][(i\omega_j +\mu)^2 - (\p+\q)^2- M^2]}\,,
\eeq
and we have again used the gap equation (\ref{eq:gap}) to eliminate $\Fones$ from the above expressions.
Note that the GL coefficients quadratic in the amplitudes could alternatively be derived as
$\alpha_2 =  \frac{1}{2}\Gamma^{-1}_\ell(0)$,   $\beta_2 =  \frac{1}{2}\Gamma^{-1}_s(0)$,
and $\alpha_{4,b} = \frac{1}{2} \frac{d}{d\q^2} \Gamma^{-1}_\ell(\q^2)|_{\q^2 = 0}$, leading
to the same results as given in section \ref{sec:GLcoeffs}. 
 
According to \Eq{eq:Omega2} the homogeneous ground state we expand about becomes unstable against developing 
inhomogeneities if $\Gamma^{-1}_\ell$ or $\Gamma^{-1}_s$ are negative in a region with nonvanishing $\q$.
Thus, as long as we are in the region where the phase diagram restricted to homogeneous condensates has a 
ground state with $\sigma_\ell = 0$, 
the second-order phase boundaries to the inhomogeneous phase  correspond to the lines in the $T$-$\mu$ plane
where one of these functions just touches the zero-axis, i.e., has a vanishing minimum at some $\q^2 \neq 0$.

\section{Numerical results}
\label{sec:num}

We now want to evaluate the GL coefficients 
and the functions $\Gamma^{-1}_\ell$ or $\Gamma^{-1}_s$ numerically
and  analyze the resulting phase structure of the model, focusing in particular on the locations of 
the critical and Lifshitz points. 
To this end we first need to fix the model parameters.
Since the NJL model is non-renormalizable this includes deciding upon a regularization scheme for the divergent momentum
integrals. 
As in our previous works, e.g., {Refs.}~\cite{Carignano:2010ac,Carignano:2018hvn,Buballa:2018hux},
we choose a Pauli-Villars (PV) inspired regularization scheme,
since a sharp three-momentum cutoff is known to lead to large regularization artifacts when dealing with inhomogeneous phases.
More specifically, we follow Ref.~\cite{Nickel:2009wj}  and regularize the vacuum contributions to the thermodynamic potential
with three PV regulators while we keep the finite medium contributions unregularized.
For the integrals $F_n(M)$ we then have
\begin{align}
       F_1(M)  & = -\frac{2 N_c}{\pi^2} \int_0^\infty dp\, p^2\, \Big[ \sum_{j=0}^3 c_j \frac{1}{E_j} 
       - \frac{1}{E}(n + \bar{n}) \Big]   \,,   \\ 
       F_2(M)  & =  \frac{N_c}{\pi^2} \int_0^\infty dp\, p^2 \, \Big[ \sum_{j=0}^3  c_j \frac{1}{E_j^3}  
       - \frac{1}{E^3}(n {+}\bar{n}) + \frac{1}{E^2}\Big(\frac{\partial n}{\partial E} + \frac{\partial \bar{n} }{\partial E} \Big)\Big] \,,
\label{eq:Fn2}
\end{align}
with $E = \sqrt{\p^2 + M^2}$, $E_j = \sqrt{E^2 + j\Lambda^2}$, the PV coefficients $c_j = \{ 1,-3,3,-1\} $, and
Fermi distribution functions $n = [\exp((E-\mu)/T) +1]^{-1}$ and $\bar{n} = [\exp((E+\mu)/T) +1]^{-1}$.
Similarly we have 
\beq
       L_2(\q^2;M) = 
       -\frac{1}{|\q|} \frac{N_c}{\pi^2}  \int_0^\infty dp\, p  \, \Big[ \sum_{j=0}^3 c_j \frac{1}{E_j}  -  \frac{1}{E} (n+\bar{n})\Big] 
       \ln\left(\frac{|\q| + 2p}{|\q| - 2p}\right) \,.
\eeq
The corresponding regulator mass $\Lambda$, the coupling constants $G$ and $K$ as well as the bare quark masses
$m_s$ can then be fitted to vacuum phenomenology. 
As a starting point we take a parameter set~\cite{DKMS}, which has been fitted to the 
vacuum values of the pion decay constant and of the masses of the pion, kaon and $\eta'$, while fixing the constituent light-quark mass
in vacuum to a value of $M_\ell^{(vac)}= 325$~MeV. This yields $\Lambda = 781.2$ MeV, $G\Lambda^2 = 4.90$, $K\Lambda^5 = 129.8$,
$m_s = 236.9$ MeV, and $m_\ell = 10.3$ MeV.
As discussed before, we work however in the light-flavor chiral limit, i.e., we set $m_\ell = 0$, instead of taking the above value. 
The vacuum constituent mass of the light quarks then takes a somewhat smaller value of about $310$~MeV, i.e., both 
$M_\ell^\mathrm{vac}$ and $\Lambda$ are roughly in the same ballpark as in similar studies in the two-flavor model.\footnote{
Typical values are $M_\ell^\mathrm{vac} = 300$~MeV and $\Lambda= 757.0$~MeV \cite{Nickel:2009wj,Carignano:2010ac}} 
Since our goals are to investigate the phase structure for both realistic and small bare strange-quark masses, we will vary $m_s$ 
in the following but refer to the fit value given above as the ``realistic'' one.

We note that the fit value of the KMT coupling in the dimensionless combination $K \Lambda^5$ is about one order of
magnitude larger than typical values using a three-momentum cutoff \cite{Rehberg:1995kh,Hatsuda:1994pi,Lutz:1992dv} . 
This is consistent with Ref.~\cite{Kohyama:2016fif}
where a systematic comparison of various regularizations schemes has been performed and where similarly large values
for $K \Lambda^5$ have been found for PV regularization.\footnote{A quantitative comparison with our parameters is not
possible since the authors of Ref.~\cite{Kohyama:2016fif} have chosen a slightly different PV scheme with
two regulators and a common regulator mass.}
Nevertheless, we will also vary the value of $K$ in order to study its effect.

\begin{center}
\begin{figure}
\includegraphics[width=.4\textwidth] {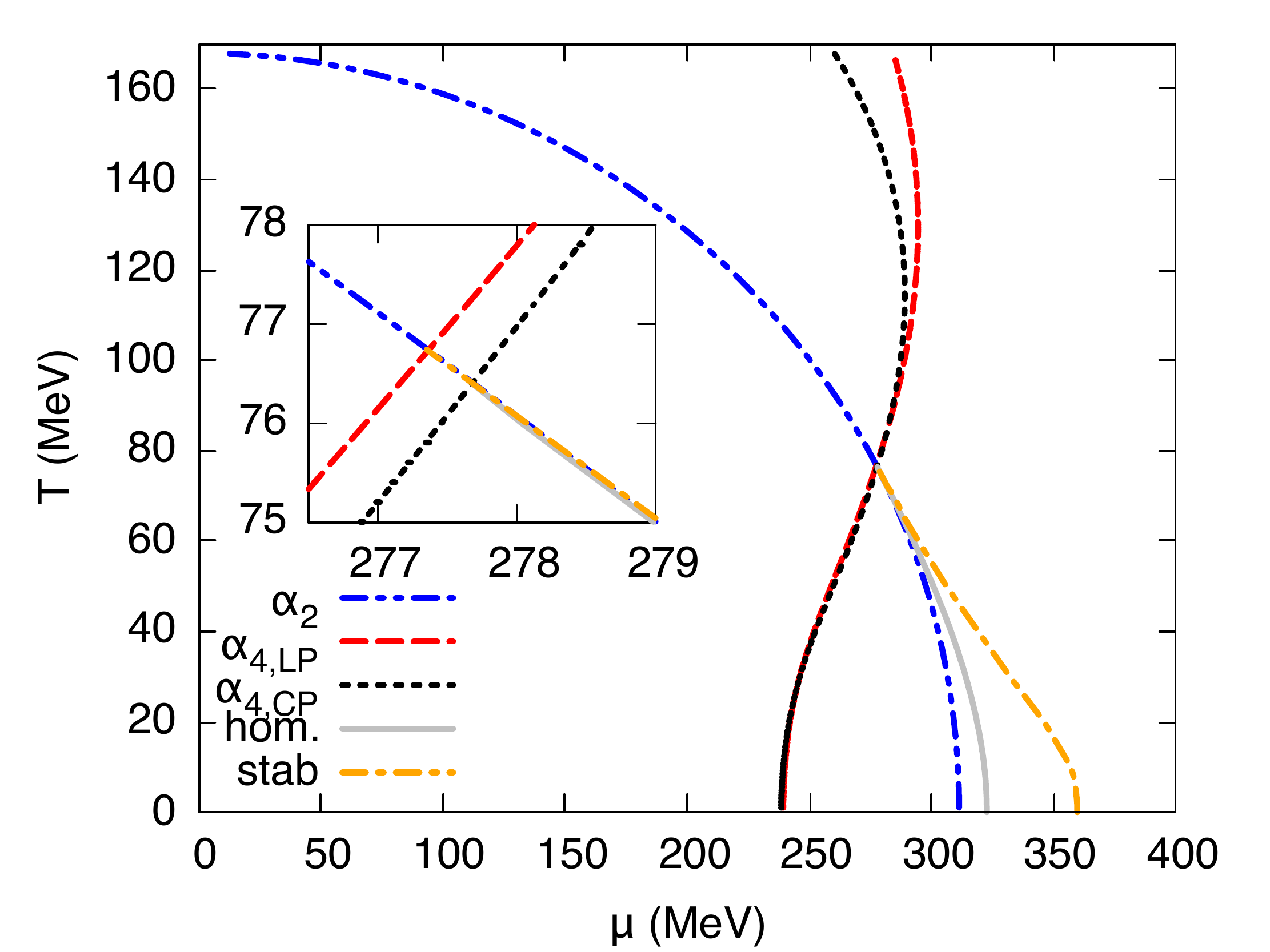}
\caption{
Phase diagram for our starting parameter set fitted to vacuum phenomenology. The inset shows
a zoomed-in view close to the CP/LP. The solid grey line denotes the first-order phase 
 boundary
for homogeneous matter, the orange dot-dashed line is the boundary of the instability region. The 
remaining lines denote the position of the roots of the GL coefficients $\alpha_2$ (blue dot-dot-dashed),
$\alpha_{4,CP}$ (black dotted) and $\alpha_{4,LP}$ (red dashed). 
Above the LP the $\alpha_2$ line corresponds to the second-order boundary between the homogeneous 
broken and two-flavor restored phase.
\label{fig:pdphys}}
\end{figure}
\end{center}

In any case, the parameters from the fit (with $m_\ell$ set to zero) turn out to be a reasonable starting point for our
studies. 
In particular, 
when we restrict the analysis to homogeneous phases, the phase boundaries, which are displayed in \Fig{fig:pdphys}, 
are
in fair agreement with typical results obtained with a three-momentum cutoff.
In the figure we also show three curves associated with the roots of the GL coefficients which, as specified in \Eq{eq:TCPLP3}, determine the locations of the CP and the LP.
With this parameter set, the two points turn out to be extremely close to each other, although, unlike in the two-flavor case, they do not coincide exactly
(see inset), with the LP being at slightly higher temperature. 

Last but not least, we show in the figure the curve where  $\Gamma^{-1}_\ell(q^2_{\rm min}) =0$, determining the onset of the instability and thus the location of the second-order phase boundary where the inhomogeneous phase meets the two-flavor chirally restored one. 
For T=0 the find that at the transition $q_{\rm min} \sim \mu \sim 350$ MeV, and as we follow this line toward higher temperatures its value gradually drops, until it reaches zero at the LP. 

It is important to note that the instability line is located at higher chemical potential than the first-order boundary of the homogeneous analysis. Therefore, coming from the two-flavor chirally restored phase and lowering the chemical potential, 
the instability toward the inhomogeneous phase occurs first and the analysis is not invalidated by a preceding homogeneous first-order transition.
In turn this means that  the first-order boundary between the homogeneous phases, including the CP, cannot be trusted.
In fact we have checked by evaluating  $\Gamma^{-1}_\ell$ that the two-flavor chirally restored solution is unstable 
 along this first-order line, indicating that the latter is entirely inside the inhomogeneous phase.

We thus find that the phase diagram is very similar to the two-flavor one, with an inhomogeneous phase covering the entire 
homogeneous first-order phase boundary. Working out explicitly 
the left phase boundary between the homogeneous broken and the inhomogeneous phase would however
require to make an ansatz for the shape of the modulation, which is outside the scope of this work.\footnote{Alternatively, the improved GL approach proposed in \cite{Carignano:2017meb} 
might also provide a good approximation for the location of the left phase transition.}
Finally we note that our numerical analysis finds no region of the phase diagram where $\Gamma^{-1}_s$ becomes negative,
i.e., the instability is driven by the light quarks only. As discussed above, because of the flavor mixing, the dynamical strange quark mass as well as the strange condensate (see \Eq{eq:Dselim}) can nevertheless be inhomogeneous as well, depending on the explicit shape of the non-strange condensates.

In order to get further insight into the nature of the splitting between the CP and the LP, 
we now investigate the effect of varying the model parameters associated with the third flavor, namely the KMT coupling $K$ and the strange current mass $m_s$ on the location of the two points. 
A first useful observation is that the equation $\alpha_{4,b}=0$ (see \Eq{eq:a4b}) does not
depend on any of these two, and therefore the location of the LP will only be affected by changes in the $\alpha_2=0$ line. 

\begin{center}
\begin{figure}
\includegraphics[width=.4\textwidth] {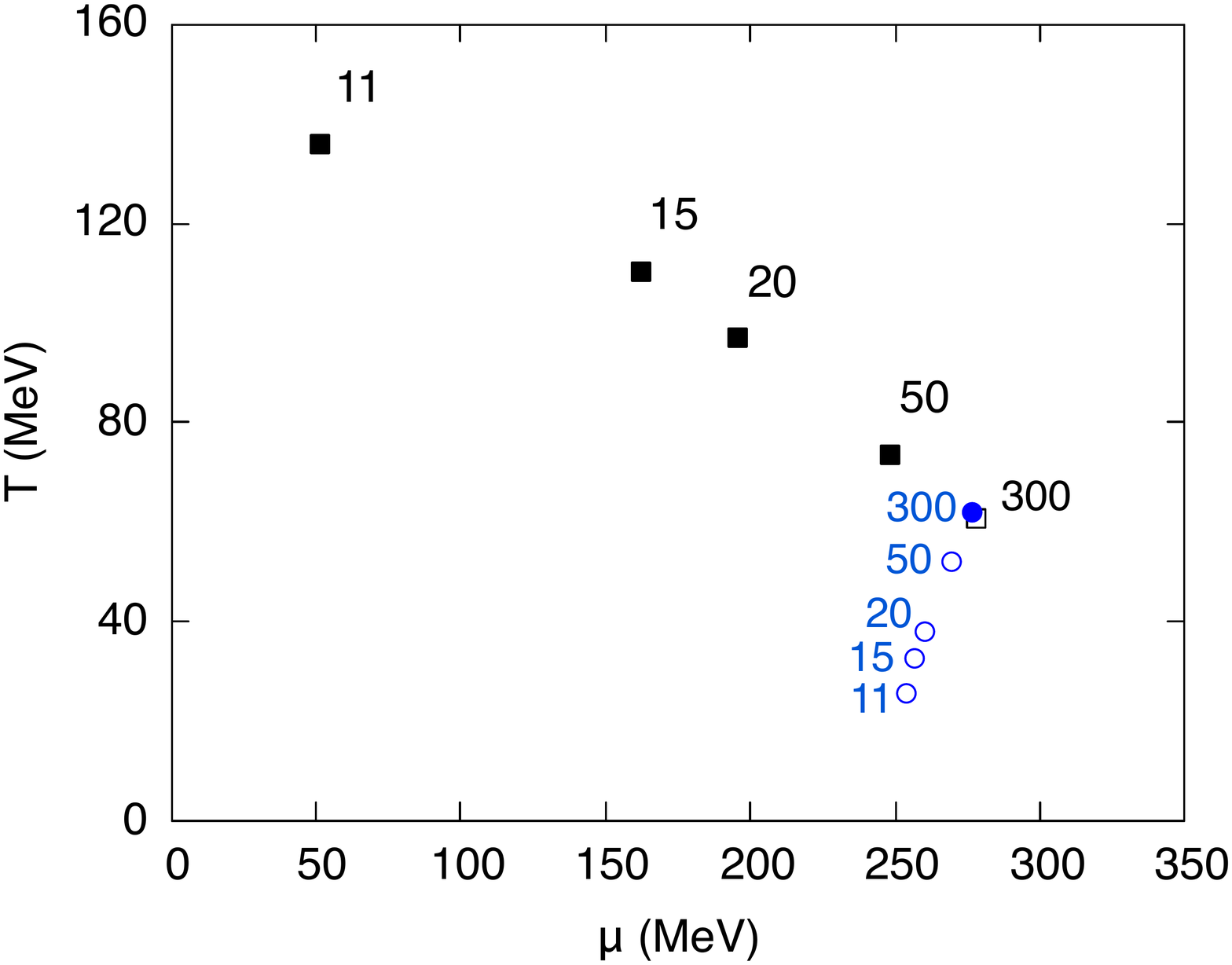}
\caption{
Locations of the CP(black squares) and LP(blue dots) for different values of $m_s$ 
(as indicated, in MeV).
Filled {(open)   symbols} denote the cases in which the GL analysis used to determine them is 
reliable (not reliable).
 \label{fig:varyms}}
\end{figure}
\end{center}

We start by showing  in \Fig{fig:varyms} the influence of the strange current mass on the location of the CP and the LP,
 varying $m_s$ from 300 MeV toward zero. 
 As can be seen from the figure, the points move rather slowly for large and moderate values of $m_s$ but speed up when the 
 current mass gets smaller. 
As expected, with decreasing $m_s$ the location of the CP moves to lower chemical potentials, until eventually it hits the temperature axis and the phase transition between the homogeneous phases becomes first order at all chemical potentials. 
With our  parameter set, this occurs for a strange current mass of around 10 MeV. This agrees well with the corresponding value of 
$m_s$ found in Ref.~\cite{Fukushima:2008}  within a three-flavor PNJL model with three-momentum cutoff, giving additional support to our parameter choice.

\begin{center}
\begin{figure}
\includegraphics[width=.4\textwidth] {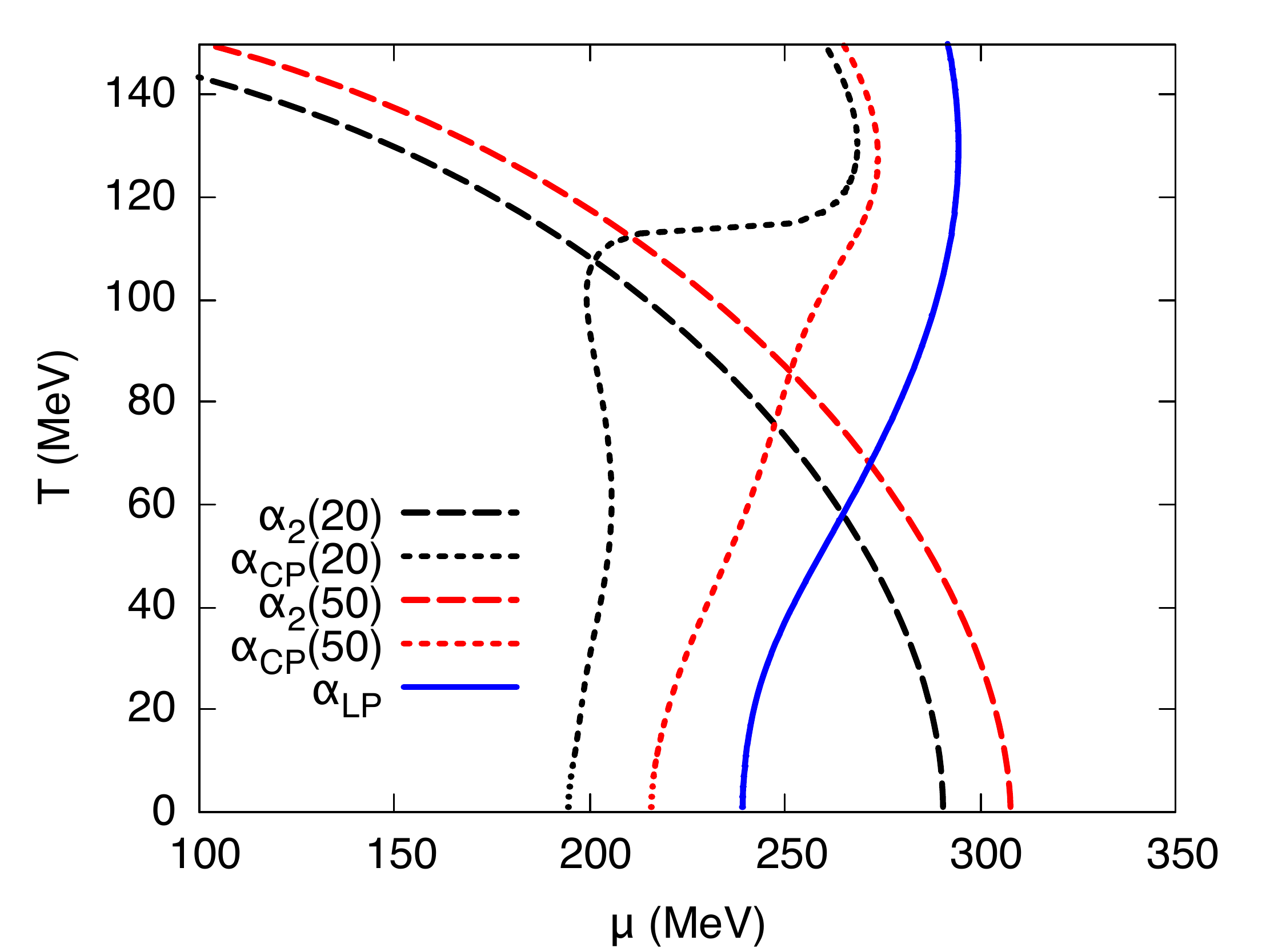}
\caption{
 Roots of GL coefficients 
 for $m_s=20$ MeV (black) and $m_s=50$ MeV (red).
The dashed lines  correspond to $\alpha_2$, the dotted ones 
to $\alpha_{4,CP}$. 
The solid blue line indicates the root of $\alpha_{4,LP}$, which is independent of $m_s$.
} \label{fig:GLms}
\end{figure}
\end{center}

The LP on the other hand follows a different trajectory, significantly splitting from the CP and moving toward lower temperatures as the current strange mass decreases.
In order to better understand this behavior, we show in \Fig{fig:GLms}  for two different values of $m_s$
the curves where the relevant GL coefficients vanish.
We see that for lower $m_s$ the $\alpha_{4,CP}$ line moves toward the temperature axis (in particular around the $\alpha_2$ line), thus explaining the corresponding movement of the CP. 
On the other hand, recalling that $\alpha_{4,LP}$ does not change at all, the relatively mild change of $\alpha_2$ 
suggests that the LP moves only little. Nevertheless, together with the shape of the $\alpha_{4,LP} = 0$ line, it explains why
the LP goes down in temperature when $m_s$ is reduced.

Again we should however pay attention to the ordering of the phase transitions. Unlike for the realistic mass case in \Fig{fig:pdphys}, 
we find that for  small values of $m_s$  the homogeneous first-order phase boundary is to the ``right'' of the inhomogeneous instability line.  In this case, coming from the two-flavor chirally restored phase and reducing the chemical potential, the homogeneous first-order transition takes place first and invalidates our analysis for the LP and the second-order phase transition to the inhomogeneous phase.
 More precisely, we find that the LPs indicated in  \Fig{fig:varyms} should not be 
trusted for $m_s \leq 50$~MeV, while
 at least part of the instability line still lies to the right of the homogeneous first-order line until $m_s \simeq 40$ MeV.
Below this value it moves completely behind it and it is then likely that the inhomogeneous phase disappears from the phase diagram.  

\begin{center}
\begin{figure}
\includegraphics[width=.4\textwidth] {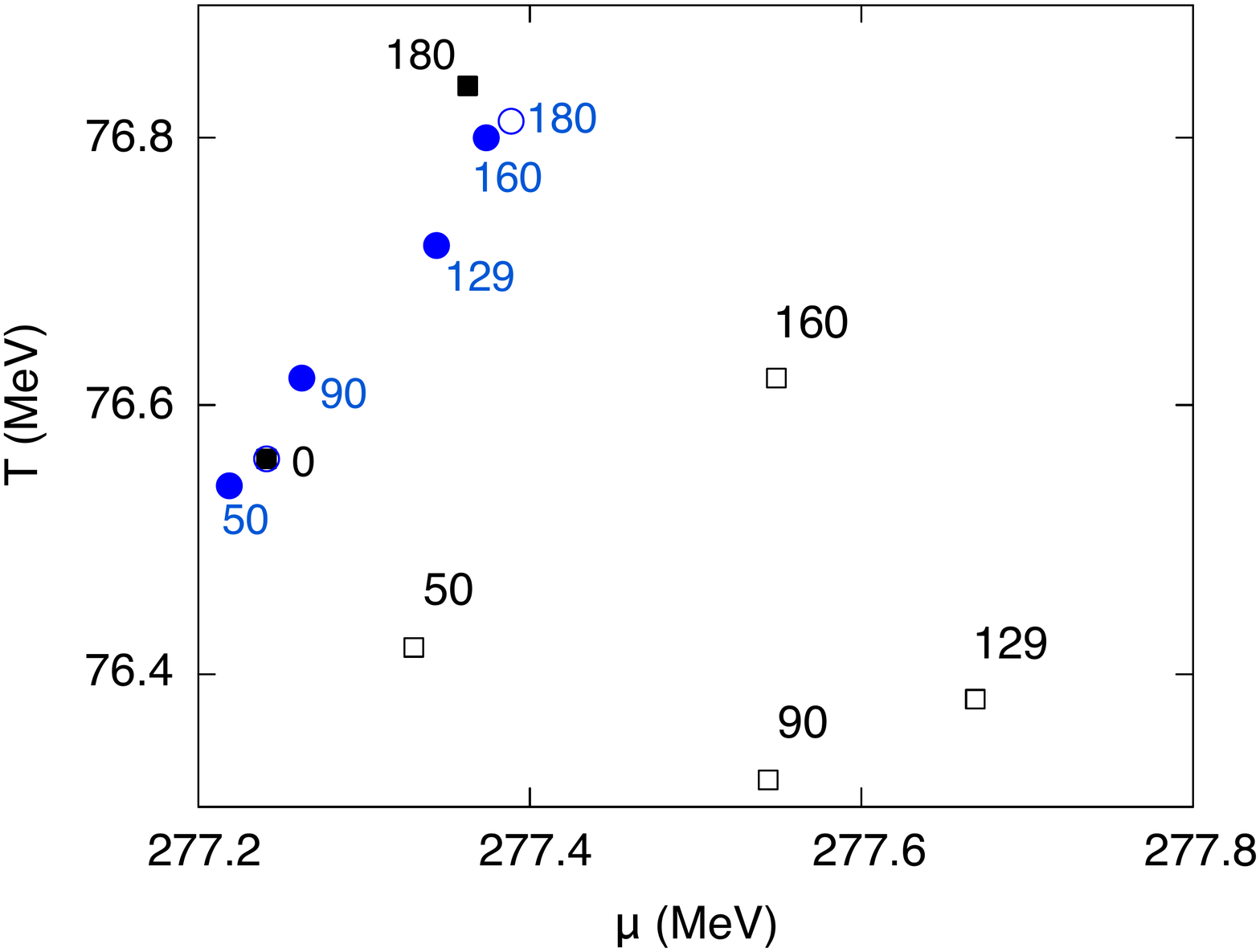}
\caption{
Locations of the CP(black squares) and LP(blue dots) for different values of $K\Lambda^5$.
For $K=0$ the two points coincide.
Filled (open) symbols denote the cases in which the GL analysis used to determine them is reliable
(not reliable).
 \label{fig:varyK}}
\end{figure}
\end{center}

In the above analysis, in order to single out the effects of varying a single parameter on the phase structure of the model,
we decided not to refit completely all the vacuum observables but simply varied $m_s$ while keeping all the other parameters fixed. 
Changing the current strange quark mass turns out to have a very small effect on the value of the vacuum constituent quark mass for light quarks\footnote{The light vacuum constituent quark mass was found to vary by less than 5 MeV throughout the whole range of $m_s$ investigated.}, whose value is known to strongly affect the phase transition in the light sector.  This was however not the case for the six-quark coupling $K$, as simply decreasing it while keeping all the other parameters fixed quickly leads to unreasonably small $M_\ell^{(vac)}$ and consequently to the disappearance of both CP and LP from the phase diagram.  In order to work around this limitation, we decided to refit the four-fermion coupling $G$ together with $K$ in order to give the same $M_\ell^{(vac)}$ as in our starting parameter set. 
The resulting locations of the CP and the LP for different values of $K$ (and refitted $G$) are shown in \Fig{fig:varyK}. 
There we can see that, after performing the refit, the locations of both points depend only very little on $K$ (note the scale!).
Within this scale, they follow again different trajectories.
As expected, the two points agree for 
 $K=0$, reproducing the known two-flavor limit as the third flavor in this case decouples completely. 
 As $K$ increases, the two split and the LP moves above the CP.
Up to $K\Lambda^5 = 160$  the situation is qualitatively similar to the one discussed in \Fig{fig:pdphys}, i.e., the
inhomogeneous phase completely covers the first-order phase boundary, invalidating the GL results for the CP.
The trend reverses if we consider extremely high values of the six-quark coupling:
For $K\Lambda^5 \gtrsim 170$ the CP reappears in the phase diagram, while the GL result for the LP can no longer be trusted.
We stress again that all these variations occur in a very limited range of temperatures and chemical potentials, as can be seen by the scale in the figure.

Finally, we would like to comment on the situation at higher densities.
From the two-flavor NJL model it is known that at higher chemical potentials a second inhomogeneous phase 
appears whose physical nature is unclear and which we termed ``inhomogeneous continent'' \cite{CB:2011}.
The corresponding phase boundary can also be determined by the stability analysis, and we find that for the three-flavor model
the inhomogeneous continent exists as well.
In fact, the existence of this phase was already reported in Ref.~\cite{Moreira:2013ura} for $T=0$. 
Moreover, these authors found that, as a consequence of the KMT interaction a third inhomogeneous phase appears
between the usual inhomogeneous ``island'' and the continent. At the lower-$\mu$ end this new phase is reached from the
two-flavor restored phase via a second-order phase transition whereas at the upper end there is a first-order transition 
back to the restored phase. 

In our calculations, i.e., with our regularization and parametrization, we do not see this intermediate inhomogeneous phase,
but we find a behavior which could be seen as precursor of this phase
and might allow us to shed some light on its nature:
Within our starting parameter set, we find that, as we increase the chemical potential beyond the second-order transition from the inhomogeneous island to the restored phase, the function $\Gamma^{-1}_\ell$ starts developing a new minimum for large values of $\bf{q}^2$, which gradually approaches zero. However, before the instability is reached, 
the partial chiral-symmetry restoration in the strange sector takes place, 
leading to a steep decrease of the dynamical strange quark mass and
pushing the minimum up away from zero.
The minimum then decreases again and eventually reaches zero at the onset of the inhomogeneous continent.

At least qualitatively, this behavior and that observed in Ref.~\cite{Moreira:2013ura} can be understood from \Eq{eq:Gammal}.
Consider a point on the second-order phase boundary between inhomogeneous and two-flavor restored phase. On this point
we have $\Gamma_\ell^{-1} = 0$, i.e., the expression in square brackets in \Eq{eq:Gammal} vanishes. From this we can see that
if we now increase the value of $\delta$, $\Gamma_\ell^{-1}$ becomes negative, i.e., the point moves inside the inhomogeneous
phase.
Hence, increasing $\delta$ increases the size of the inhomogeneous regime, moving the upper phase boundary of the 
inhomogeneous island  upwards and the lower boundary of the continent downwards in $\mu$.
The situation is complicated however by the fact that $\delta$ depends on $M_{s,0}$ and thus indirectly on $T$ and $\mu$.
As long as $M_{s,0}$ is roughly constant,  we have $\delta\propto K/G^2$, i.e., increasing $K$ (and simultaneously decreasing $G$
by the parameter refit) enhances the value of $\delta$ and thus of the inhomogeneous region. In contrast, if we keep
$K$ fixed but increase the value of $\mu$, the strong decrease of $M_{s,0}$ 
 related to the partial chiral-symmetry restoration in the strange sector 
lowers $\delta$ considerably,
in our case preventing an early onset of the continent. If on the other hand the  decrease of $M_{s,0}$ occurs after the second inhomogeneous phase has already started, the related  decrease of $\delta$ can shut this phase off again, and the continent reappears only at even higher values of $\mu$. It is possible that this is what happened in  Ref.~\cite{Moreira:2013ura}, and it would be interesting to check whether this scenario is indeed what is occurring within the model regularization and parametrization used in that work.

\section{Conclusions}
\label{sec:conclusions}

In this work we investigated within the NJL model
 how
the formation of inhomogeneous chiral condensates 
 is affected by
the inclusion of strange quarks, 
which 
 are coupled to
 the light flavors via a 
  KMT determinant interaction,
 mimicking the effects of the axial anomaly.
The use of a Ginzburg-Landau analysis allowed us to infer the location of the critical
and the Lifshitz points in a general way without having to specify a 
 shape
for the spatial dependence of the chiral condensate.
 Furthermore, using a complementary stability analysis,
we were able to determine 
the location of the second-order phase transition where inhomogeneous condensates become favored over chirally restored matter.
 The latter also indicates that it is favorable only for the light quark condensates to become inhomogeneous, whereas no instability was found in the strange quark channel.
This does not exclude that an inhomogeneous strange quark condensate is induced by the coupling to the light quarks,
 but this depends on the favored shape of the spatial modulations.

Our first main result is that the axial anomaly leads to a splitting between the CP of the phase transition for homogeneous matter and the LP associated with the inhomogeneous phase.
This splitting turns
out generally to be very small, with the exception of the limit of 
very small strange current quark masses, for which the CP moves toward the temperature axis and 
eventually
disappears from the phase diagram, whereas the LP does not follow the same behavior.  For a realistic parameter set fitted to vacuum phenomenology, the two points are nevertheless found to almost 
 coincide for a wide range of values of the six-quark coupling $K$. 
 It is important however to stress that the points do not coincide exactly, and for a wide range of parameters explored in this work 
 we find that the first-order chiral phase transition which is typically present when restricting the model 
 analysis to homogeneous matter is completely covered by the inhomogeneous phase. 
  
  A closer inspection of the behavior of the GL coefficients allowed us to better understand the behavior of the two points and the magnitude of their splitting,
 while our complementary stability analysis let us determine in several cases the location of the phase boundary where the inhomogeneous phase terminates. 
    Consistently with the results of {Ref.}~\cite{Moreira:2013ura}, we find an enhancement of the inhomogeneous phase due to the coupling with strange quarks, although, as already mentioned, 
    the effect turns out to be quite small, and we  did not
    witness the appearance of any additional inhomogeneous window in the phase diagram compared with the two-flavor case.

We recall that our study relies on the approximation of small amplitudes (and additionally of small gradients for the GL analysis), so it can only provide a reliable result if we are able to expand about 
 a suitable homogeneous solution, which for simplicity we assumed to be the 
two-flavor restored phase. 
It would be tedious but straightforward to extend the method to expand about massive homogeneous solutions in the 
light-quark sector as well, as we have done in Ref.~\cite{Buballa:2018hux} for the two-flavor model away from the chiral limit.
We note that in this case we could even study the stability of the homogeneous chirally broken phase against inhomogeneities,
while this region was not accessible to our present analysis.  
But even then we stay restricted to second-order phase transitions. 
It would therefore be desirable to perform a full numerical evaluation of the model thermodynamic potential in order to obtain the 
complete phase structure of the model. 
This would require the introduction of specific Ans\"atze on the spatial dependence of the chiral condensate and a full numerical diagonalization of the quark Hamiltonian, as done, e.g., in {Ref.}~\cite{Carignano:2012sx}.  
In principle one could also perform an Ansatz-free minimization of the thermodynamic potential within lattice field theory,
which has already been applied to analyze inhomogeneous phases in lower-dimensional models  \cite{Wagner:2007he,Heinz:2015lua,Pannullo:2019prx,Winstel:2019zfn}
but obviously becomes computationally extremely expensive in $3+1$ dimensions.
Another way to go beyond the small amplitude constraint would be to perform 
an extended GL analysis, as done in \cite{Carignano:2017meb}, although the determination of higher order GL coefficients would become significantly more involved due to the additional contributions in the model Lagrangian.

Ultimately we are of course interested in the phase structure of QCD, i.e., one should go beyond effective models and the mean-field approximation. 
In this context it is interesting to note that in Ref.~\cite{Fu:2019hdw} hints for an inhomogeneous phase have been found
within an FRG study of the QCD phase diagram for $2+1$ flavors. It was found that  in a certain regime in the vicinity and
above the (homogeneous) chiral phase boundary the mesonic wave-function renormalization constant at vanishing momentum 
$Z_\phi(0)$ becomes negative. 
As the authors point out, this could indicate an instability toward a spatially modulated regime but the signal is not fully conclusive.
In fact, $Z_\phi(0)$ basically corresponds to the GL coefficient $\alpha_{4,b}$ in our language, so finding a negative value is
a necessary but not a sufficient requirement for an inhomogeneous phase. 
This becomes clear if we recall that $\alpha_{4,b}$ is proportional to $\frac{d}{d\q^2} \Gamma^{-1}_\ell(\q^2)|_{\q^2 = 0}$.
Hence, a negative value hints at a minimum of $\Gamma^{-1}_\ell$ at nonzero $\q^2$ but does not tell whether its value at the
minimum is positive or negative. 
Indeed, in a previous FRG study within the quark-meson model \cite{Tripolt:2017zgc}
it was found that the static pion two-point function develops a maximum (corresponding to a minimum of 
$\Gamma^{-1}_\ell$) at nonzero $\q^2$ but does not change its sign, so that no instability occurred.

In Ref.~\cite{Fu:2019hdw} the region with negative $Z_\phi(0)$ starts at a chemical potential well below the CP.
Hence, if it was indeed related to an instability, it would mean that the CP and most likely the entire first-order phase boundary are covered by an inhomogeneous phase. 
Unfortunately, a fully conclusive analysis of this issue is extremely involved due to several competing order parameters.
In this sense, the analysis performed in our work can act as a guide to give a qualitative picture of the phase structure of dense 
$2+1$-flavor  quark matter as a starting point for these more refined studies to begin with.

\subsection*{Acknowledgments}
We thank Dominic Kraatz and Michel Stillger for providing us with their model parameters.	
We acknowledge support by the Deutsche Forschungsgemeinschaft (DFG, German Research Foundation) through the CRC-TR 211 ``Strong-interaction matter under extreme conditions" - project number 315477589 - TRR 211.
S.C. has been supported by the projects FPA2016-81114-P and FPA2016-76005-C2-1-P (Spain), and by the project 2017-SGR-929 (Catalonia).

\end{document}